\documentclass[aoas,MSNbibl,nameyear,seceqn,dvips]{arximspdf}
\usepackage{graphicx}

\doi{10.1214/11-AOAS483}
\volume{5}
\issue{2B}
\pubyear{2011}
\firstpage{1207}
\lastpage{1228}

\makeatletter
\renewcommand{\epsilon}{\varepsilon}

\def\bptnote#1{}

\makeatother

\begin{document}
\begin{frontmatter}

\title{State-space solutions to the dynamic magnetoencephalography
inverse problem using high performance computing\thanksref{T1}}
\runtitle{Dynamic state-space methods for MEG}

\begin{aug}
\author[A]{\fnms{Christopher J.} \snm{Long}\corref{}\ead[label=e1]{cjlong1@gmail.com}\tsup{2,3}},
\author[B]{\fnms{Patrick L.} \snm{Purdon}\tsup{4,5}},
\author[C]{\fnms{Simona}~\snm{Temereanca}\tsup{4}},
\author[D]{\fnms{Neil U.} \snm{Desai}\tsup{6}},
\author[E]{\fnms{Matti~S.}~\snm{H\"{a}m\"{a}l\"{a}inen}\tsup{4,7}}
\and
\author[F]{\fnms{Emery N.} \snm{Brown}\ead[label=e2]{enbrown1@mit.edu}\tsup{4,5,6,7}}
\runauthor{C. J. Long et al.}
\pdfauthor{Christopher J. Long,
Patrick L. Purdon,
Simona Temereanca,
Neil U. Desai,
Matti~S. Hamalainen,
Emery N. Brown}
\affiliation{\tsup{2}Imperial College London, \tsup{3}GlaxoSmithKline,
\tsup{4}Harvard Medical School, \tsup{5}Massachusetts General Hospital,
\tsup{6}Massachusetts Institute of Technology (MIT),
\tsup{7}Harvard/MIT Division of Health Sciences \& Technology}
\address[A]{C. J. Long\\
GlaxoSmithKline Clinical\\ \quad Imaging Center\\
Hammersmith Hospital\\
London W12 0NN\\
UK\\
\printead{e1}} 
\address[B]{P. L. Purdon\\
Athinoula A. Martinos Center\\ \quad for Biomedical Imaging\\
Massachusetts General Hospital\hspace*{28pt}\\
Harvard Medical School\\
Charlestown, Massachusetts\\
USA}
\address[C]{S. Temereanca\\
Athinoula A. Martinos Center\\ \quad for Biomedical Imaging\\
Massachusetts General Hospital\\
Harvard Medical School\\
Charlestown, Massachusetts\\
USA}
\address[D]{N. U. Desai\\
Department of Electrical Engineering\\ \quad \& Computer Science\\
Massachusetts Institute of Technology\\
Cambridge, Massachusetts\\
USA}
\address[E]{M. S. H\"{a}m\"{a}l\"{a}inen\\
Athinoula A. Martinos Center\\ \quad for Biomedical Imaging\\
Massachusetts General Hospital\\
Harvard Medical School\\
Charlestown, Massachusetts\\
USA\\
and\\
Harvard/MIT Division of\\ \quad  Health Sciences \& Technology\\
Massachusetts Institute of Technology\\
Cambridge, Massachusetts\\
USA}
\address[F]{E. N. Brown\\
Department of Brain\\ \quad And Cognitive Sciences\\
Massachusetts Institute of Technology\\
Cambridge, Massachusetts\\
USA\\
and\\
Department of Anesthesia\\ \quad \& Critical Care\\
Massachusetts General Hospital\\
Boston, Massachusetts\\
USA\\
and\\
Harvard/MIT Division\\\quad of Health Sciences \& Technology\\
Massachusetts Institute of Technology\\
Cambridge, Massachusetts\\
USA\\
\printead{e2}}
\end{aug}
\thankstext{T1}{Supported in part by the National Science Foundation
through TeraGrid resources
provided by the Texas Advanced Computing Center (TACC).
This work was also supported by NIH Grants NIBIB R01 EB0522 (E.N.B.),
DP2-OD006454 (P.L.P.),
K25-NS05780 (P.L.P.), DP1-OD003646 (E.N.B.), NCRR P41-RR14075,
 R01-EB006385 (E.N.B., P.L.P., M.S.H.) at
Massachusetts General Hospital.}

\received{\smonth{9} \syear{2010}}
\revised{\smonth{4} \syear{2011}}

%
\begin{abstract}
Determining the magnitude and location of neural sources within the
brain that are responsible for generating magnetoencephalography (MEG)
signals measured on the surface of the head is a challenging problem in
functional neuroimaging. The number of potential sources within the
brain exceeds by an order of magnitude the number of recording sites.
As a consequence, the estimates for the magnitude and location of the
neural sources will be ill-conditioned because of the underdetermined
nature of the problem. One well-known technique designed to address
this imbalance is the minimum norm estimator (MNE). This approach
imposes an $L^2$ regularization constraint that serves to stabilize and
condition the source parameter estimates. However, these classes of
regularizer are static in time and do not consider the temporal
constraints inherent to the biophysics of the MEG experiment. In this
paper we propose a dynamic state-space model that accounts for both
spatial and temporal correlations within and across candidate
intracortical sources. In our model, the observation model is derived
from the steady-state solution to Maxwell's equations while the latent
model representing neural dynamics is given by a random walk process.
We show that the Kalman filter (KF) and the Kalman smoother [also known
as the fixed-interval smoother (FIS)] may be used to solve the ensuing
high-dimensional state-estimation problem. Using a well-known
relationship between Bayesian estimation and Kalman filtering, we show
that the MNE estimates carry a significant zero bias. Calculating these
high-dimensional state estimates is a computationally challenging task
that requires High Performance Computing (HPC) resources. To this end,
we employ the NSF Teragrid Supercomputing Network to compute the source
estimates. We\vadjust{\goodbreak} demonstrate improvement in performance of the state-space
algorithm relative to MNE in analyses of simulated and actual
somatosensory MEG experiments. Our findings establish the benefits of
high-dimensional state-space modeling as an effective means to solve
the MEG source localization problem.\vspace*{3pt}
\end{abstract}

%

\end{frontmatter}

\section{Introduction}
Electromagnetic source imaging is a neuroimaging technique that permits
study of neural events on a millisecond timescale. This type of imaging
reveals brain dynamics that cannot be seen with imaging modalities such
as functional magnetic resonance imaging (fMRI) or positron emission
tomography (PET) that capture brain activity on a much slower
timescale. Two of the most important examples of electromagnetic
imaging are electroencephalography (EEG) and magnetoencepholography
imaging (MEG). These modalities acquire multiple time series recorded
at locations exterior to the skull and are generated, respectively, by
electric and magnetic fields of neuronal currents within the cortex of
the brain [\citet{r1}, \citet{r2}]. In these experiments subjects execute a task
that putatively activates multiple brain areas. In the case of EEG
recordings, the mode of acquisition involves affixing an array of
electrodes to the surface of the scalp and measuring the potential
differences relative to a~reference node. In contrast, MEG records
extremely weak magnetic fields emanating from the brain on the order of
femtoTesla, that is, $10^{-15}$ Tesla [\citet{r2}]. To put these
quantities into context, the magnetic field of the earth is around
$5\times10^{-5}$ Tesla while that of a present-day magnetic resonance
scanner used for medical diagnostic imaging ranges between $1.5$ and
$7$ Tesla. Because the magnetic fields of the brain are extremely weak,
a specially shielded room and highly sensitive array of detectors,
called superconducting quantum interference devices (SQUIDs), are
required to undertake the MEG recordings.\looseness=1

Generating images of the brain's cortical activity from MEG time-series
recordings requires solving a high-dimensional ill-posed inverse
problem. In the case of MEG imaging, the nonunique nature of the
inverse problem arises because the number of potential
three-dimensional brain sources (as a consequence of the biophysics of
the problem) is infinite, while the measured two-dimensional SQUID
array typically contain 306 magnetometer and gradiometer detectors.

The accuracy of the inverse solution depends critically on two main
features of the problem, specifically, the precision of the forward
model and the choice of inverse solution algorithm. The MEG forward
model is designed to describe how activity propagates from sources in
the cortex to the SQUID detectors. These forward models are normally
constructed by combining information about head and brain geometry with
the electromagnetic properties of the skull, dura and scalp to yield a
numerical approximation to Maxwell's quasistatic equations; see, for
example, \citet{r1}. Under this forward model the strength of the
magnetic field recorded at a~given SQUID detector near the surface of
the head is approximately proportional to the reciprocal of its squared
distance from a given cortical source. The second consideration we face
relates to the choice of source localization algorithm. Solving this
localization problem is one of the most technically challenging in MEG
imaging research and has been an active area of methods development for
both EEG and MEG over the past two decades. Historically, two of the
most popular methods for solving the EEG/MEG inverse problem have been
the dipole-based and (linear) distributed-source methods.

Dipole-based methods model the underlying neuronal sources as discrete
current dipoles whose location, orientation and amplitude are unknown~%
quan\-tities to be estimated [\citet{r4}, Uutela, H{\"a}m{\"a}l{\"a}inen and
  Salmelin \citeyear{r5}]. Choosing the number of
dipoles to include in the model is problematic. Eigenvalue
decomposition methods have been developed to address this model
selection issue [\citet{r4}]. However, these approaches still require
subjective interpretation when the eigenvalue distribution does not
yield an obvious distinction between the signal and noise subspaces. In
addition, finding the best-fitting parameters for the multidipole model
requires nonlinear optimization. Since the measured fields depend
nonlinearly on the dipole position parameters, typical minimization
routines may not yield the globally optimal estimates for these
parameters. Proposed solutions to this problem include the MUSIC
(MUltiple SIgnal Classification) algorithm [\citet{r6}], global
optimization algorithms and partly heuristic model constructions that
use interactive software tools [\citet{r5}]. However, none of these
methods utilize the temporal sequence of the signals, that is, if the
original data points are first permuted and an inverse permutation is
applied to the source estimates, the results are the same as without
permutation. Finally, in many important clinical and neuroscientific
applications, such as epilepsy, sleep or general anesthesia, the dipole
model is inappropriate, since the generating currents are most likely
distributed across large areas on the cortex.\looseness=-1

In the distributed source models, each location in the brain represents
a~possible site of a current source. Since the number of unknown
sources vastly outnumbers the number of EEG or MEG data recordings,
constraints are required to obtain a unique solution. The minimum-norm
estimator (MNE) employs an $L^{2}$-norm data ``fidelity'' term to
quantify the relationship between the observed magnetic field
recordings and the estimated source predictions and an $L^{2}$-norm
regularization or penalty term on the magnitude of those solutions.
While the minimum-norm estimate (MNE) yields the solution with the
smallest energy (sum of squares) across the solution space [\citet{r1}],
it tends to consistently favor low-amplitude solutions that are located
close to the scalp surface. The relative contribution of the data term
and source term can be controlled or tuned through a regularization
parameter. In MNE, the data term is specifically the $L^{2}$-norm of
spatially whitened data while the source term is the $L^{2}$-norm of
the current amplitudes. A well-known variant of this approach is the
LORETA (LOw-Resolution Electromagnetic Tomography Algorithm), where the
$L^{2}$-norm\vadjust{\goodbreak} source term specified in MNE is replaced by a weighted
norm that approximates the spatial Laplacian of the currents [\citet{r7}].
In the case of minimum-current estimates (MCE) [\citet{r8}], the source
term is taken to be the $L^{1}$-norm of the source current amplitudes.
A consequence of this choice is that the estimates will be sparse as
opposed to diffuse, leading to families of solutions that may closely
resemble those uncovered through a dipolar analysis.\looseness=-1

Beamformer approaches, originally inspired by problems in radar and~%
so\-nar array signal processing [\citet{r33}], seek to localize EEG/\break MEG
activity by specifying a series of spatial filters, each tuned to
maximize sensitivity to a particular spatial location within the brain
[\citet{r9}]. These methods assume, however, that EEG/MEG sources are
spatially uncorrelated, and are limited by interference crosstalk from
sources outside the focus of the beamformer spatial filter. A more
recent related approach by   \citet{r10} features a graphical
model framework to characterize the probabilistic relationships among
latent and observable temporal sources for averaged ERP data. Once a
low-dimensional statistical data subspace is estimated, the algorithm
computes full posterior distributions of the parameters and
hyperparameters with the primary objective of alleviating temporal
correlation between spatially distinct sources (source separation). In
later work [\citet{r11}] this approach is extended somewhat to include
explicit temporal modeling by drawing on estimated weightings from a
predefined family of smooth basis functions.   \citet{r12} explores
an alternate procedure to empirical Bayesian source modeling that
attempts to simultaneously capture and model correlated source-space
dynamics in the presence of unknown dipole orientation and background
interference.

Bayesian approaches to MEG source localization have included \citet
{r13}, \citet{r14} and \citet{r15}. Collectively,~the\-se techniques
construct the source localization problem within an instantaneous
empirical Bayes framework. These approaches specify a nested and
hierarchical series of spatial linear models to be subsequently solved
under a penalized Restricted Maximum Likelihood (ReML) procedure. The
multiresolutional nature of this regime carries with it a measure of
spatial adaptation for the estimated hyperparameters since the spatial
clusters of activated dipoles are amalgamated across several
(orthogonal) scales. At the same time the most likely set of spatial
priors for each competing model is chosen using information selective
model ranking procedures. The latter approach [\citet{r15}] extends the
use of ReML to engender a~framework aiming to move from parametric
spatial priors to larger sets of imputed sources with localized but not
necessarily continuous support. The principle advantage of this
extension is its intrinsic ability to capture source solutions ranging
from sparse dipole models at one end of the localization spectrum to
dense but smooth distributed configurations of spatial dipoles at the
other.\vadjust{\goodbreak}

State-space methods featuring
latent temporal models have recently been proposed. For example,
studies by \citet{r16}  and \citet{r17} have used a random-walk dynamical
model with Laplacian spatial constraints to represent the dynamics of
EEG source currents, employing the Kalman filter and recursive
least-squares framework to perform source localization. \citet{r18} used
a state-space model to capture latent neuronal dynamics by placing a
first-order autorgressive (AR) model within a full spatiotemporal
variational Bayes procedure. Each ensemble of sources is characterized
by establishing an effective region of influence in which the hidden
neural state-dynamics are assumed constant. This leads to a temporal
representation of each region in terms of a single average dynamic.

\citet{r19} have developed a mixed $L^{1}L^{2}$-norm
penalized model to estimate MEG/EEG sources. This methodology allows
for spatially sparse solutions while ensuring physiologically plausible
and numerically stable time course reconstructions through the use of
basis functions. Since estimation of this kind of mixed-norm
regularized problems represents a computational challenge, convex-cone
optimization methods are implemented to efficiently compute the MEG/EEG
source estimates. These procedures are shown to yield significant
improvement over conventional MNE approaches in both simulated and
actual MEG data.

In the majority of MEG stimulus response experiments, the objective is
to estimate the activity in a given brain region based on multiple
repetitions of the same stimulus. The time interval of the MEG
recordings after the onset of the stimulus is usually around 1,000
msec. Consequently, the optimal estimate of the source activity at a
given time point should be based on all the data recorded in this time
window, as opposed to only the recordings within a small instantaneous
subinterval as is the case in the MNE algorithm.

Thus, in the current work, we model the spatiotemporal structure of the
MEG source localization problem as a full-rank state-space estimation
procedure. In contrast to previous related approaches, for example,
\citet{r16}, \citet{r17} and \citet{r19}, we recover the underlying neural state dynamics
across the entire solution space using all information available in the
MEG measurements to compute a new source estimate at each location and
time point. We do not restrict our state-covariance structure to
operate either through reduced-rank approximations or through
``small-volume'' state-space models that operate within localized spatial
neighborhoods [\citet{r16}, \citet{r17}]. The analysis computes activity at a
particular source with activity at that source combined with
(potentially all) neighboring sources in both preceding and subsequent
time periods (spatiotemporal correlation). We find that solution of the
resulting high-dimensional state-space model requires the application
of high performance computing (HPC) resources.

We show that two solutions that are naturally suited to this type of
dynamic inverse problem are the Kalman filter and the fixed-interval
smoother [\citet{r22}, \citet{r34}]. These algorithms are based on a~series of
mathematical relationships that employ principles from electromagnetic
field theory to relate the observed MEG measurements to the underlying
dynamics of the current source dipoles. In contrast to popular source
localization methods such as MNE where the regularization constraints
are formulated in terms of a fixed-error covariance matrix, the
state-space estimators feature time-varying error covariances and
propagate past (and future) information into the current update.

We describe how these algorithms may be used to perform this state
estimation by conducting the computation on the NSF Teragrid
Supercomputing Network. In Section \ref{SC_implementation} we detail
the necessary supercomputing methods needed to implement the
high-dimensional state-space estimation. We demonstrate the improvement
in performance of the state-space algorithm relative to MNE in analyses
of a simulated MEG experiment and actual somatosensory MEG experiment.

\section{Theory}

We assume that in an MEG experiment a stimulus is applied~$T$ times and
at each time for a (short) window of time following the application of
the stimulus MEG activity is measured simultaneously in $S$ SQUID
recording sites. Let $Y_{s,r} (k )$ denote the measurement at
time $k$ at~lo\-cation $s$ on repetition $r$ where $k=1,\ldots,N,
s=1,\ldots,S$ and $r=1,\ldots,T$. We take $Y_s(k)=T^{-1}\sum
^{T}_{r=1}Y_{s,r}(k)$ and we define $Y_k=Y (Y_1(k),\ldots
,Y_S(k) )$ to be the $S\times1$ vector of measurements recorded at
time $k$. We assume that there are $P$ current sources in the brain and
that the relation between the MEG recordings and the sources is defined by
\begin{equation} \label{eqn1}
Y_k=HJ_k+\epsilon_k,
\end{equation}
where $J_k=(J_{k,1},\ldots,J_{k,P})$ is the $3P\times1$ vector of
source activities at time~$k$, each $J_{k,i}$ is a $3\times1$ vector,
and $H$ is the $S\times3P$ lead field matrix, derived as described in
\citet{r1}. This matrix is approximated in practice using the known
conductivity profile of an MRI-derived T1 image to derive a one-layer
head model using the boundary element method [\citet{r20}]. Furthermore,
$\epsilon_k$ is a zero mean Gaussian noise with covariance matrix
$\Sigma_\epsilon$. To build on an idea originally suggested in the EEG
literature [\citet{r16}, \citet{r17}], we assume that the $J_k$ behave like a
stochastic first-order process with a potentially weighted neighborhood
(six direct neighbors in the case below) constraint defined at voxel
$p$ by
\begin{equation} \label{eqn2}
J_{k,p}=J_{k-1,p}+\frac{1}{6}m^{-1}_\ell\sum_{\ell\in{M{(\ell
)}}}J_{k-1,\ell}+v_k,
\end{equation}
where $M(\ell)$ is some small neighborhood around voxel $p$ and $m_{\ell
}$ is a mask of (estimated) autoregressive weights, and $v_{k}$ is a
$3P\times1$ vector of Gaussian\vadjust{\goodbreak} noise with mean zero and covariance
matrix $\Sigma_w$. The linear structure in equation (\ref{eqn2}) means
that we can generalize this structure into matrix format such that all
sources in the brain (i.e., not only local relationships) are
encompassed in the model. That is,
\begin{equation} \label{eqn3}
J_k=FJ_{k-1}+v_k.
\end{equation}
If $F$ is an identity matrix, the state dynamics reduce to a random
walk process. Equations (\ref{eqn1}) through (\ref{eqn3}) define the
observation and latent time-series equations, respectively, for a
state-space model formulation that can be used to provide a dynamic
description of the MEG source localization problem.

\subsection{Standard MEG inverse solution}

The standard approach to the MEG source localization problem is to
solve an $L^2$ regularized least-squares problem at each time $k$. That is,
\begin{equation}\label{eqn4}
 \|Y_k-HJ_k \|^2_{\Sigma_{\epsilon}}+ \|J_k-\mu \|^2_C,
\end{equation}
where $ \|y-x \|^2_Q$ denotes the Mahalanobis distance between
the vectors $y$ and $x$ with error covariance matrix $Q$, $\mu$ is an
offset parameter (prior mean) and $C$ is the regularization covariance
matrix [\citet{r21}].

If we take $\mu=0$ and define the source covariance matrix as $C=\lambda
{R}$, where $R$ is a diagonal, scaled matrix normally computed in
advance, and if $\lambda>0$, then at each time $k$ an instantaneous MEG
source estimate (the MNE solution) is given as
\begin{equation} \label{eqn5}
J^{\mathrm{MNE}}_k=\lambda{RH^T} (\lambda{\mathit{HRH}^T}+\Sigma_\epsilon )^{-1}Y_k
\end{equation}
for $k=1,\ldots,N$. The MNE estimate is a local Bayes' estimate because
it uses in its computation only the data $Y_k$ at time $k$ and the
Gaussian prior distribution with mean $\mu=0$ and covariance matrix
$\lambda{R}$. Hence, the MNE estimation procedure imposes no temporal
constraint on the sequence of solutions.

\subsection{Kalman filter solution}

Given the state-space formulation of the MEG observation process in
equations (\ref{eqn1}) and (\ref{eqn3}), it follows that the optimal
estimate of the current source at time $k$ using the data $Y_1,\ldots
,Y_N$ up through time $N$ is given by the Kalman filter [\citet{r22}]
%
\begin{eqnarray}
\label{eqn6}
J_{k|k-1}&=&FJ_{k-1|k-1}, \\
\label{eqn7}
W_{k|k-1}&=&W_{k-1|k-1}+\Sigma_w, \\
\label{eqn8}
K_k&=&W_{k|k-1}H^T [HW_{k|k-1}H^T+\Sigma_\epsilon ]^{-1}, \\
\label{eqn9}
J_{k|k}&=&W_{k|k-1}H^T [HW_{k|k-1}H^T+\Sigma_\epsilon
]^{-1}Y_k \nonumber
\\[-8pt]
\\[-8pt]
&&{}+ [I-K_kH ]J_{k|k-1} ,
\nonumber
\end{eqnarray}
which simplifies to
\begin{eqnarray}
\label{eqn10}
J_{k|k}&=&J_{k|k-1}+K_k [Y_k-HJ_{k|k-1} ], \\
W_{k|k}&=& [I-K_kH ]W_{k|k-1}
\end{eqnarray}
for $k=1,\ldots,N$, given initial conditions $J_0\sim N
(0,W_{0|0} ), \Sigma_w=W_{0|0}$. At each time $k$ the Kalman
filter computes $p (x_k|Y_1,\ldots,Y_k )$, which is the
Gaussian distribution with mean $J_{k|k}$ and covariance matrix
$W_{k|k}$. In terms of the regularization criterion function in
equation (\ref{eqn4}), the Kalman filter solution is equivalent to
choosing at time $k$, $\mu= [I-K_kH ]FJ_{k-1|k-1}$ and
$C=W_{k|k-1}$, where we have used the matrix inversion lemma to
re-express the Kalman filter update in equation (\ref{eqn9}) in order
that we can compare it directly with the MNE solution in equation (\ref
{eqn5}). From this comparison we see that the Kalman solution improves
upon the MNE solution in two important ways. First, in the Kalman
solution, the offset parameter and the regularization matrix are
different at each time $k$ and are given, respectively, by $\mu$ and the
one-step prediction error covariance matrix $W_{k|k-1}$. The MNE
solution at each time $k$ has a fixed prior mean $\mu=0$ and
(temporally) fixed regularization matrix $C=\lambda{R}$. Because of
this choice of regularization constraint, equation (\ref{eqn9}) shows
that the Kalman filter estimate at time $k$ is a linear combination of
$J_{k-1|k-1}$, the current source estimate at time $k-1$, and $Y_k$,
the observations at time $k$. In contrast, the MNE solution at each
time $k$ is a~linear combination of the observed data and a fixed prior
mean of $\mu=0$. In this regard, the MNE estimate biases the solution
at each time $k$ toward~0. Taken together, these observations show that
the stochastic continuity assumption in the Kalman state model results
in a time-varying constraint upon the fluctuating source estimates.

\subsection{Fixed-interval smoothing algorithm solution}
Because the MEG time series are often fixed length recordings, we can
go beyond the estimate $J_{k|k}$ provided by the Kalman update and
compute the posterior density $p (x_k|Y_1,\ldots,Y_N )$ at
time $k$ given all the data in the experiment. To do so, we combine the
Kalman filter with the fixed-interval smoothing algorithm (FIS) [\citet
{r22}, \citet{r34}], that may be computed as follows:
%
\begin{eqnarray}
\label{eqn11}
A_k&=&W_{k|k}W^{-1}_{k+1|k}, \\
\label{eqn12}
J_{k|N}&=&x_{k|k}+A_k [J_{k|k}-J_{k+1|k} ], \\
\label{eqn13}
W_{k|N}&=&W_{k|k}+A [W_{k|N}-W_{k+1|k} ]A^T_k
\end{eqnarray}
for $k=N-1,\ldots,1$ and initial conditions $J_{N|N}$ and $W_{N|N}$
computed from the last step of the Kalman filter $N$. It is well known
that $p (x_k|Y_1,\ldots,Y_N )$ is a Gaussian distribution with
mean $J_k|N$ and covariance matrix $W_{k|N}$ [\citet{r22}, \citet{r34}].

\subsection{Practical considerations}
Implementation of the Kalman filter first requires making estimates of
both the initial state, the state covariance matrix $\Sigma_w$ and the
error covariance matrix $\Sigma_\epsilon$. We estimated the noise
covariance matrix $\Sigma_\epsilon$ from measurements taken in the
scanner in the absence of a~subject. We next estimated the initial
state as the MNE solution $J^{\mathrm{MNE}}_0$ and the state covariance matrix
by first computing the MNE source estimates $J^{\mathrm{MNE}}_1,\ldots
,J^{\mathrm{MNE}}_N$, subsequently deriving $\Sigma_w$ from the sample
covariance matrix using a differenced sequence of these static estimates.

\section{Supercomputer implementation}
\label{SC_implementation}
Historically the Kalman filter has found widespread use in several
high-dimensional modeling domains, including weather forecasting [\citet
{r29}] and oceanography [\citet{r30}, \citet{r31}]. Kalman filters and
fixed-interval smoothers are advantageous in these scenarios as, under
assumptions of linearity and normality, they are (near) optimal
estimators. In addition, their desirable properties hold across a wide
variety of time-varying linear (and nonlinear) models. However, in its
standard form the Kalman filter is computationally prohibitive for
these classes of problems. In the example applications listed above,
the numerical calculations are often carried out on systems of state
dimension $N\sim O(10^7)$ with state covariance matrices of size
$N^2\sim O(10^{14})$ [\citet{r29}]. Computationally, the most intense
aspects of the Kalman algorithm stem from the prediction update in the
covariance matrices that require costly linear algebraic updates at
each time step. Since the dynamical error structure of these systems is
often well understood, many numerical solutions to these kinds of
paradigms, for example, \citet{r29} and \citet{r31}, employ a range of
model-reduction techniques in their formulations to achieve
computational tractability.

In the case of the MEG and EEG inverse problem, the solution space is
${\sim} O(10^3-10^4)$, leading to error covariances of dimension $P^2\sim
O(10^6\mbox{--}10^8)$. Thus, the computational problem is not quite so
intensive as in the forecasting applications, meaning that we can
feasibly employ full-rank state-estimation methods to compute their
solution on an HPC system. When performing Kalman filtering, the
computations dominating each time step involve three high-dimensional
full-rank matrix multiplications, leading to a~total approximate
computational cost of $3P^3$ (excluding auxiliary lower dimensional
linear operations). When taking into account ancillary variables, the
amount of memory required is at least 24$+$ gigabytes for each update
operation. In addition, the FIS requires one additional such
multiplication at each time step followed by a full-rank $P \times P$
matrix inversion ($2P^3$). Also, FIS requires approximately three
gigabytes of storage space per time point in order to save the
prediction and update covariances (in 32-bit storage format) estimated
during the forward pass of the Kalman filter. The scale of these
computations renders them infeasible on nearly any state-of-the-art
standalone computing resource. To address this limitation, we
parallellized the Kalman and FIS filtering computations such that the
data-intensive parts of the algorithm were distributed across multiple
nodes of a~High Performance Computing system [\citet{r32}]. For this
purpose we utilized the NSF Teragrid resource at the TACC (Texas
Advanced Computing Center). This resource comprised a 1024-processor
Cray/Dell Xeon-based Linux cluster with a total of 6.4 Teraflops
computing capacity.\looseness=-1

\section{Results}

\subsection{Simulated MEG experiment}

We designed a set of simulation studies to compare the performance of
the dynamic localization methods against MNE. Prior to constructing the
dynamic simulation, we first computed the spatial conductance profile
across the head and specified the spatial resolution of the discretized
solution space (source locations). To restrict this source space to the
cortical surface, we employed anatomical MRI data obtained from a
single subject with a high-resolution T1-weighted 3D sequence
(TR/TE/flip${}={}$2\mbox{,}530 ms/3.49~ms/$7^{\circ}$, partition thickness${}={}$1.33~mm, matrix${}
= 256\times256\times128$, field of view${}={}$21~cm$\times$21~cm) in a 3-T
MRI scanner (Siemens Medical Solutions, Erlangen, Germany). The
geometry of the gray--white matter surface was subsequently computed
using an automatic segmentation algorithm to yield a triangulated model
with approximately 340,000 vertices [\citet{r23}]. Finally, we utilized
the topology of a recursively subdivided icosahedron with approximately
5~mm spacing between the source nodes to give a cortical sampling of
approximately $10^4$ locations across both hemispheres.

Using MNE, we chose as the region of interest a section of the left
hemisphere over the primary somatosensory and motor cortices. We
computed a single layer homogeneous forward model or lead field matrix
$H$ over all the sampled voxels in the left and right hemispheres. The
region of interest contained 125 active voxels from the more than
10,000 gray matter voxels that could be potential sources for an
observed magnetic field under this parcellation. For this simulation,
therefore, the number of active sources was $P_{\mathrm{active}}=125$, and the
number of measurement channels was $S=306$. We next constructed $H$
with dimension $306\times5\mbox{,}120$ (left-hemisphere), corresponding to a
spatial sampling of 5 mm. Note that we chose to estimate only the
dominant normal MEG component at each vertex (i.e., $z$-direction) as
opposed to estimating the triplet of $x,y $ and $ z$ contributions from
which both magnitude and directional information could be resolved. In
each of the chosen 125 voxels in the source space, we generated an
``activation'' signal consisting of a mixture of 10 and 20 Hz frequencies
modulated by a~0.4 Hz envelope to simulate typical signal patterns
commonly observed in the motor and somatosensory cortices; see, for
example, \citet{r24}. Next we added Gaussian noise to all $P=5\mbox{,}120$
vertices to generate an image-wide signal-to-noise ratio (SNR) that
ranged between 0.1 and 2 (in line with typical SNRs encountered in MEG
studies). Specifically, we defined SNR as
%
\begin{equation}
\label{eqn14}
\frac{ \|HJ^2 \|}{SP\sigma^2}.
\end{equation}
When computing the source reconstructions we set the observation
covariance matrix as the empty room covariance, that is, that which is
generated from background noise measurements taken from the system
while the subject is absent from the scanner. We computed three inverse
solutions: the MNE solution, the KF solution, and the FIS solution. In
this simulation study, we recovered the source estimate for each method
and for each value of SNR using three different choices of tuning
parameter ($\lambda=0.5, 1, 3$). The time-course length was 120 time
points and with an assumed sampling rate of 600 Hz, equated to a data
segment of about 200 msec.

To examine the benefits of the dynamic state-space procedures in
relation to the MNE source localization algorithm, we computed the
Kalman and Fixed-Interval Smoothing filters, generating estimates of
the source locations and amplitudes within the entire cortex. When
applied to the 200~ms window of MEG data, this analysis averaged about
one hour for each simulated record when distributed across 16 of the
CRAY-DELL cores. For each choice of SNR (through a sweep of 20 SNRs
choices equispaced between 0.1 and 2), the FIS algorithm took around
two hours on 24 CRAY-DELL computational nodes, leading to a total CPU
time of around 1\mbox{,}280 hours for the whole simulation. Storage of the
prediction and update covariances for each SNR required approximately
110 gigabytes of disk space, resulting in a total of 2~Terabytes of
storage for each of the three choices of tuning parameter ($\lambda$),
that is, 6 Terabytes in total.

\begin{figure}
\vspace*{-3pt}
\includegraphics{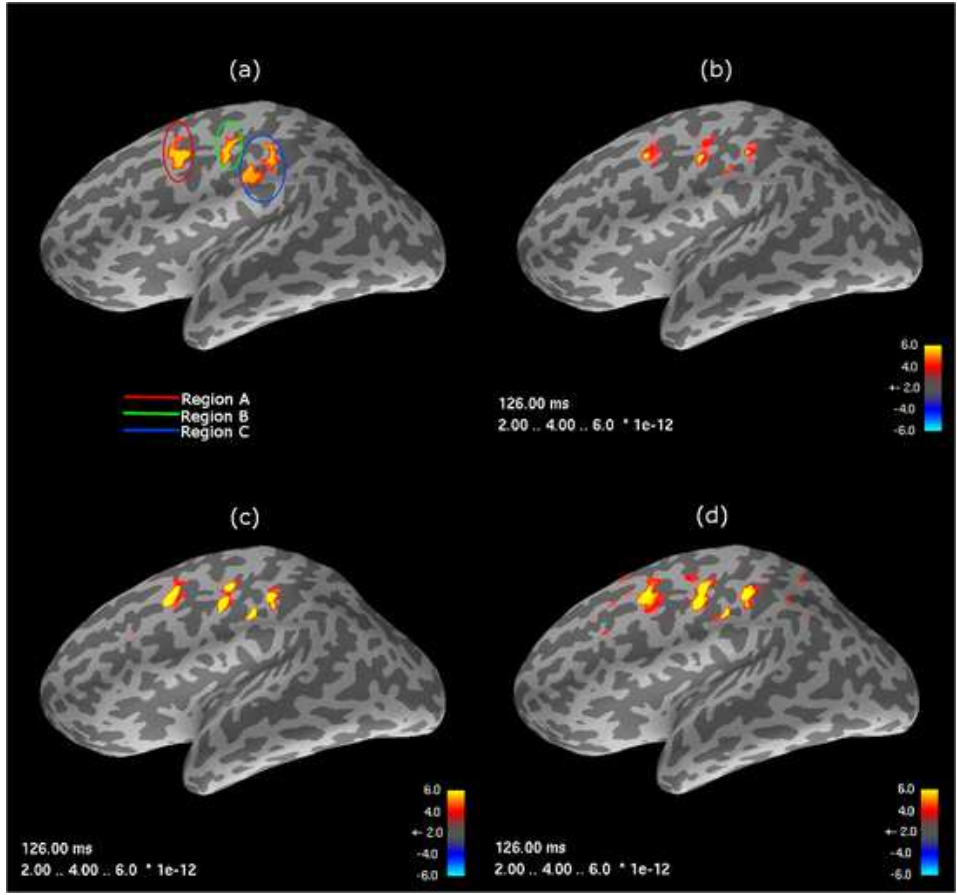}

\caption{Activation maps in three regions, A, B and C at the peak
response time of 126 ms for: \textup{(a)}~True signal; and source estimates
computed by \textup{(b)} MNE; \textup{(c)} the Kalman filter; and \textup{(d)} the Fixed Interval
Smoother. These images were computed using a regularization parameter
($\lambda=1$).}
\vspace*{-5pt}
\label{figure1}
\end{figure}

\begin{figure}

\includegraphics{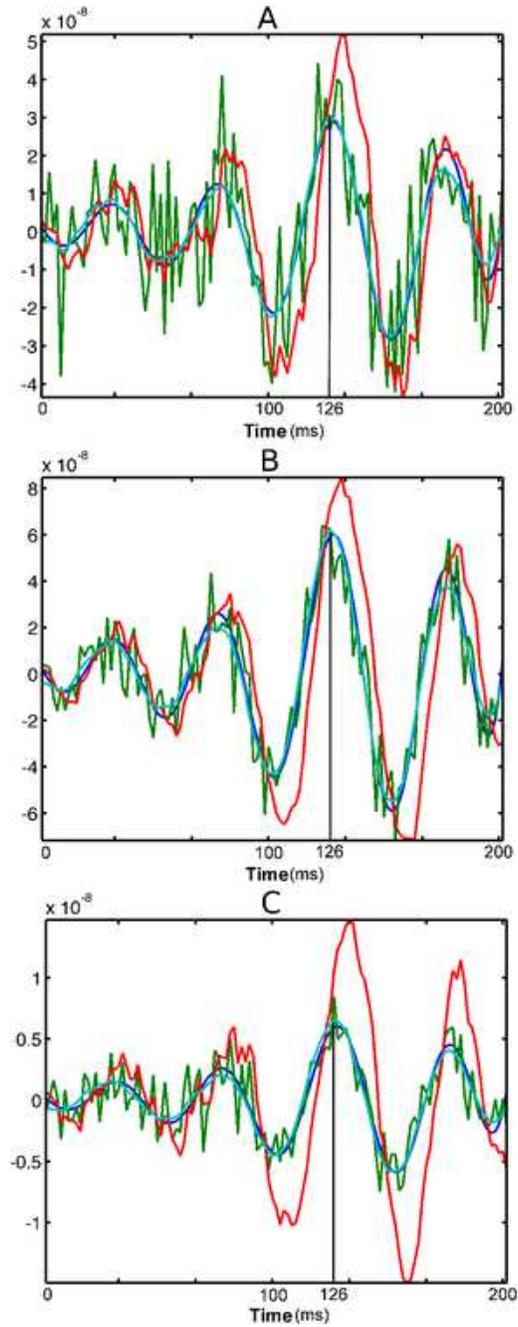}

\caption{Time course of source estimates from the simulated MEG
experiment for the maximally responding voxel relative to the MNE
solution, where each panel (from top to bottom) corresponds directly to
each of the three regions, A, B and C denoted in Figure~\protect\ref{figure1}\textup{(a)}.
Green denotes the MNE solution, red the KF, dark blue is the FIS
solution, while cyan represents the true signal.}
\label{figure2}
\end{figure}
\begin{figure}

\includegraphics{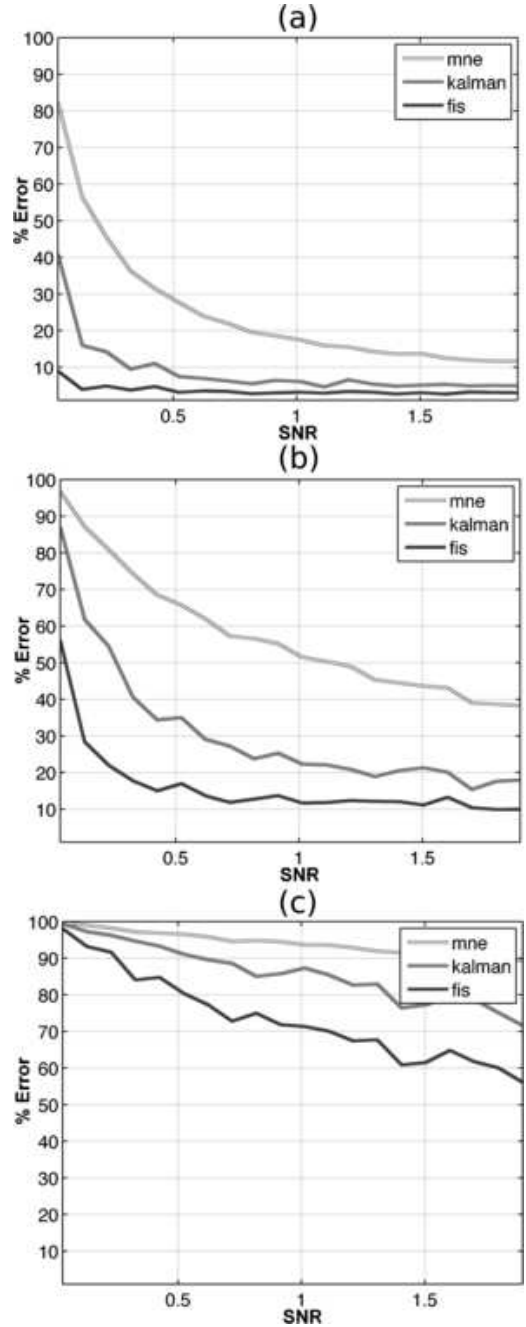}

\caption{Average percentage relative change in MSE computed as a
function of SNR from the simulated MEG experiment for all voxels on the
cortical surface for each localization method: MNE (light gray), Kalman
filter (dark gray), and Fixed Interval Smoother (charcoal). \textup{(a)} $\lambda
=0.5,$ \textup{(b)}~$\lambda=1,$ \textup{(c)}~$\lambda=3$.}
\label{figure3}
\end{figure}

\begin{figure}

\includegraphics{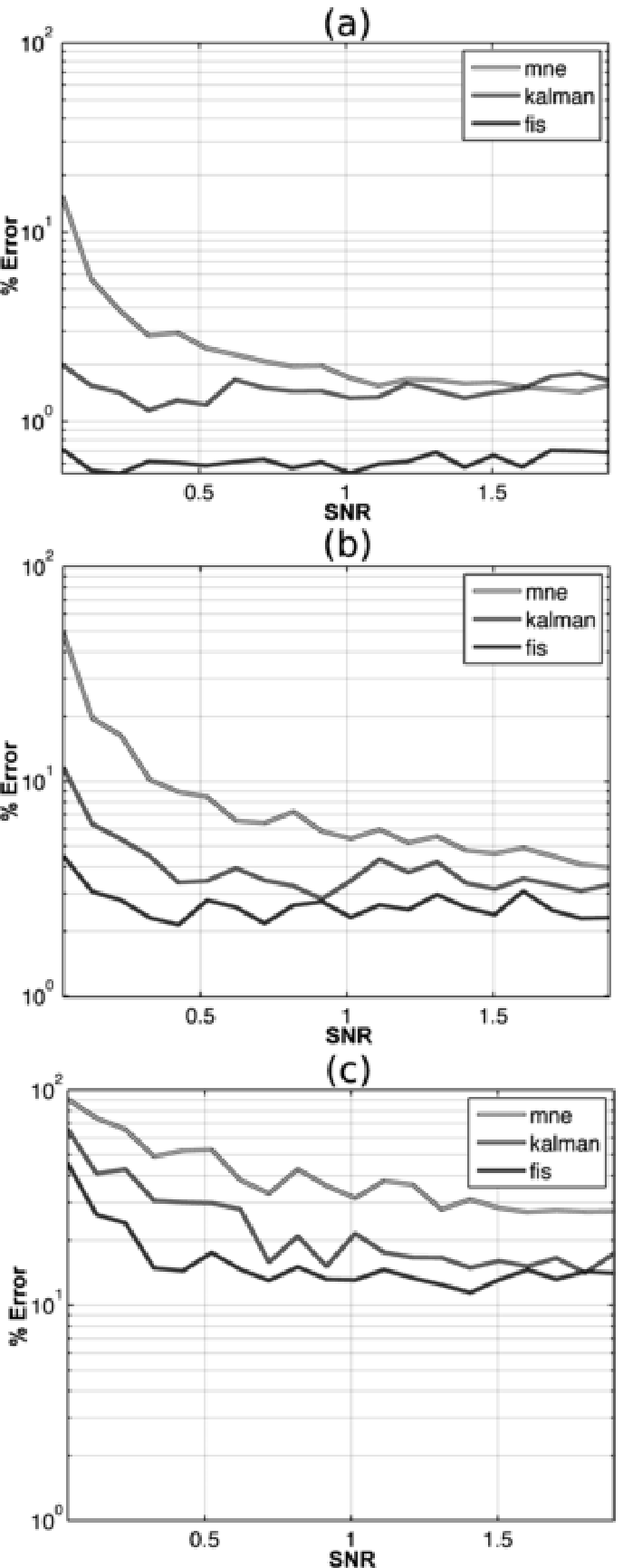}

\caption{Average percentage relative change in MSE computed as a
function of SNR in the simulated MEG experiment across all \textup{active} sources for each localization method: MNE (light gray), Kalman
filter (dark gray), and Fixed Interval Smoother (charcoal). \textup{(a)} $\lambda
=0.5$, \textup{(b)} $\lambda=$ \textup{(c)}, $\lambda=3$.}
\label{figure4}
\end{figure}

Figure \ref{figure1} illustrates the spatial extent of the simulated
signal (a) taken as a snapshot at the peak of the damped sinusoidal
signal. Yellow to red color map values indicate progressively smaller
current sources ($J$) representing ground truth, panel~(a), or the
reconstructed estimates (b)--(d). Figure \ref{figure1} panels (b)--(d) show
the estimates as computed with (b) the minimum norm estimate, (c) the
Kalman filter and (d) the Fixed Interval Smoother. For each method, the
SNR as defined by equation (\ref{eqn14}) and the tuning parameter
governing the strength of the estimate smoothness were set to
$\mathit{SNR}=\lambda=1$.

Figure \ref{figure2} depicts temporal reconstructions for the three
methods with the choice of $\mathit{SNR}=\lambda=1$. For each region A, B or C,
the time series at the voxel corresponding to the peak MNE response was
plotted and overlaid onto the ``ground-truth'' simulated signal. The
state-space reconstructions show considerably less variability than the
static MNE time course, however, the KF approach displays elevated bias
as compared to the FIS method. Furthermore, the KF method contains a
temporal shift in relation to the simulated signal that is not apparent
in MNE and FIS. These fundamental differences in the temporal behavior
of the three methods are consistent in space, across SNR, and are
invariant to choice of $\lambda$.

Figures \ref{figure3} and \ref{figure4} illustrate the behavior of MSE
(Mean-Squared Error) as a function of SNR for the three choices of
tuning parameter: panel (a) $\lambda=0.5$, panel (b) $\lambda=1$, and
panel (c) $\lambda=3$ between the true signal and the estimated signal
for each estimation method. In Figure \ref{figure3} MSE is calculated
at all vertices in the reconstructed cortical surface and averaged
spatially for each choice of SNR. The fixed-interval smoother shows the
lowest MSE which indicates that the estimated source curves are closer
to the ground truth simulations compared to the alternative methods. By
contrast, the MNE approximations show the least favorable recovery of
the simulated sources. In Figure \ref{figure4} we computed the MSE at
each vertex for each source reconstruction method in the same manner
but calculated the average MSE only within the three activated regions.
These plots, therefore, examine the behavior of the source
reconstructions by zooming into areas that contain signal, thus
decoupling the model fits to areas of true signal from those vertices
containing background noise. The three methods show similar behavior as
in Figure \ref{figure3} in the activated regions, with the
fixed-interval smoother again showing superior performance to the other
two methods. As a consequence of the relative MSE errors exhibiting
significantly smaller values in these signal-rich regions, we display
their effects on a log linear scale to better delineate the salient
differences between methods.

\section{Somatosensory MEG experiment}
To illustrate the state-space and MNE methods on an actual MEG
experiment, we estimated the sources of the alpha (8--13 Hz) and mu (10
Hz and 20 Hz) spontaneous brain \mbox{activity}\break \mbox{using} anatomically-constrained
whole-head MEG. Synchronous cortical\break rhythmic activity of large
populations of neurons is associated with distinct brain states and has
been the subject of extensive investigation using unit
electrophysiology, as well as EEG, MEG and fMRI techniques [\citet
{r24}, \citet{r25}]. The alpha rhythm is recorded over the posterior parts of the
brain, being strongest when the subject has closed eyes and reduced
when the subject has open eyes. Sources of alpha rhythm have been
identified with MEG and EEG to lie in the parietooccipital sulcus and
calcarine fissure [\citet{r26}]. The mu rhythm represents activity close
to 10 and 20 Hz in the somatomotor cortex, and is known to reflect
resting states characterized by lack of movement and somatosensory
input. The 20 Hz component tends to be localized in the motor cortex,
anterior to location of the 10 Hz component [\citet{r24}], being
specifically suppressed or elevated at topographic locations
representing moving or resting individual limbs, respectively. By
comparison, the 10 Hz component arises in the somatosensory cortex, and
is thought to reflect the absence of sensory input from the upper limbs
[\citet{r27}, \citet{r28}].

Following approval by the Massachusetts General Hospital Human
Research Committee, the MEG signals were acquired in a healthy male
subject in a magnetically and electrically shielded room at the
Martinos Center for Biomedical Imaging at the Massachusetts General
Hospital. MEG signals were recorded from the entire head using a
306-channel dc-SQUID Neuromag Vectorview system (Elekta Neuromag,
Elekta Oy, Helsinki, Finland) while the subject sat with his head
inside the helmet-shaped dewar containing the sensors. The magnetic
fields were recorded simultaneously at 102~lo\-cations, each with 2
planar gradiometers and 1 magnetometer at a~sampling rate of 600 Hz,
minimally filtered to (0.1--200 Hz). The positions of the electrodes in
addition to fiduciary points, such as the nose, nasion and preauricular
points, were digitized using the 3Space Isotrak II System for
subsequent precise co-registration with MRI images. The position of the
head with respect to the helium-cooled dewar containing the measurement
SQUIDS was determined by digitizing the positions of four coils that
were attached to the head. These four coils are subsequently employed
by the MEG sensors to assist in co-registration.

Ongoing magnetic activity was thus recorded under the following three
conditions: (1) at rest with eyes closed; (2) at rest with eyes open; and
(3) during sustained finger movement with eyes open. As described in
previous studies, examination of raw data revealed strong alpha
activity when the subject had eyes closed, and dampened or no alpha
rhythm when the subject had eyes open (i.e., at rest or during
sustained finger movement conditions). The mu activity was strongest at
rest (i.e., with eyes open or closed) and suppressed during sustained
fingers movement condition. 2 s long periods of raw data with either
large amplitude alpha waves (e.g., during eyes-closed conditions) or
large amplitude mu waves (e.g., during hand/finger resting conditions)
were used as the input signal for MEG source localization. In addition,
the empty MEG room noise was recorded just before the start of the
experiment and subsequently used as measurement noise in the estimation
of the alpha and mu activity generators.

\begin{figure}

\includegraphics{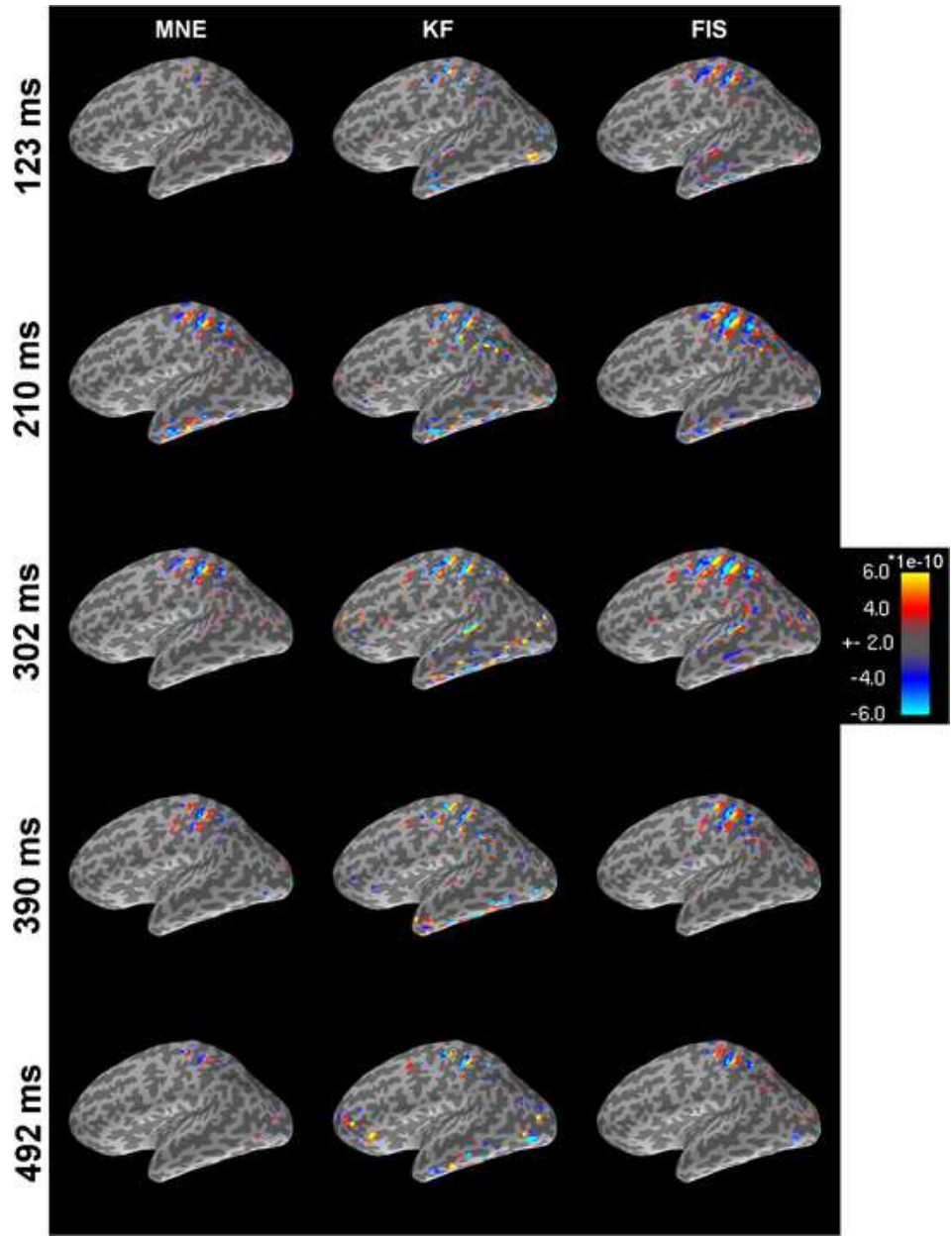}

\caption{Source activation maps showing instantaneous reconstructions
for the sensorimotor task. Each column shows the temporal evolution at
the five peak modes of the mu-rhythm for each of the three methods:
Minimal Norm Estimator (MNE); Kalman filter (KF); and Fixed Interval
Smoother (FIS). The red and blue color maps correspond to inward and
outward current flow, respectively. This reconstruction was carried out
using a regularization parameter value of $\lambda=1$.}
\label{figure5}
\end{figure}

\begin{figure}

\includegraphics{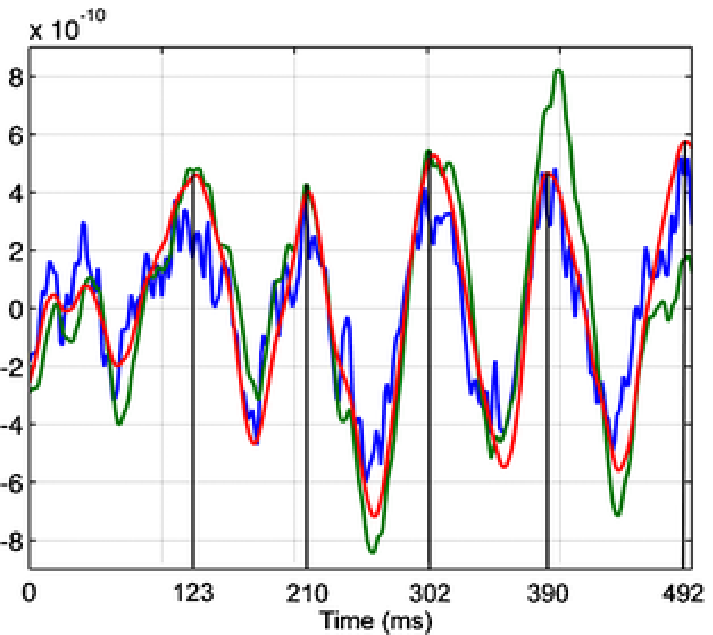}

\caption{Representative time course reconstructions extracted from
locations in the somatosensory cortex at 492 ms in Figure \protect\ref
{figure5}: MNE (blue), Kalman filter (green) and Fixed Interval
Smoother (red).}
\label{figure6}
\end{figure}

Figure \ref{figure5} depicts the net estimated current distributions
characterized in each panel by the length of each current triplet. In
the inverse calculations, the orientation of the estimated currents was
fixed such that they lay perpendicular to the cortical mesh. Each panel
reveals the relative effect of the state-space approaches as compared
to those gained from the static MNE technique as the dynamics of the
mu-rhythm time course unfold (Figure \ref{figure6}). In all cases, we
can observe regions in the central sulcus showing strong activation,
but the FIS and KF solutions show stronger activations than those
obtained through the MNE method. Moreover, Figure \ref{figure6} shows
that the reconstructed time courses of the KF and FIS solutions are
much less variable than those of the MNE solution.

In summary, analysis of the somatosensory experiment showed approximate
agreement between MNE and the state-space models. As in the simulated
MEG experiment, we note that the magnitudes of the estimate sources
were greater for the state-space models (Figure \ref{figure5}). The
temporal dynamics of the FIS agreed more closely with the MNE than with
the KF (Figure \ref{figure5}). Not surprisingly, because the
state-space models imposed a temporal constraint, their estimates were
smoother than the MNE estimates (Figure \ref{figure6}). Also, as in the
simulated MEG experiment, the KF estimates of the source amplitudes
were larger than those computed by either MNE or the FIS.

\section{Discussion}
We have developed and implemented a state-space model solution to the
MEG inverse problem. We examined its performance relative to the MNE
algorithm on simulated and actual MEG experimental data. A~simple
analytic analysis showed that the state-space approach offers two
important improvements over MNE. Namely, the KF/FIS estimate equation
(\ref{eqn9}) uses a dynamic $L^2$-norm regularization mean equation
(\ref{eqn6}) and covariance matrix equation (\ref{eqn7}) \mbox{update} at each
step, whereas the MNE estimate equation (\ref{eqn5}) uses a static
$L^2$-norm regularization zero mean and static covariance matrix to
compute its instantaneous estimate. Making the prior mean zero at each
update biases every solution toward zero. This makes it difficult to
identify sources that are nonzero but weak (low amplitude). In
contrast, the KF/FIS estimate is more likely to identify a weak source
because if the source estimate at the previous time point, that is, 1
msec before, was similarly weak, it would be combined with the current
observation to identify what would most likely be an attenuated source
at the current time point. Second, the KF source estimates use all of
the observations from the start of the experiment to the current time,
then the FIS equation (\ref{eqn12}) uses the KF source estimates to
compute new estimates based on all of the experimental observations. In
contrast, the MNE source estimate uses only the observations recorded
at the current time. Given that the MEG updates are computed every
millisecond, the solutions at adjacent time periods are likely to be
strongly correlated. Our state-space approach imposes a dynamic
$L^2$-norm regularization constraint to use this temporal information,
whereas MNE imposes the same static $L^2$-norm regularization
constraint at each time point and thus does not consider the temporal structure.
Our simulation studies demonstrated that the KF and FIS estimates are
more accurate in terms of MSE (Figures \ref{figure3} and \ref{figure4})
compared with MNE. Although the spatial maps of the state-space and MNE
had comparable spatial extent, the KF and FIS estimates captured more
accurately the magnitude of the activations (Figure \ref{figure1}).
This accounts for the lower MSE for the FIS and the KF relative to MNE
despite small numbers of erroneous activations in the case of the
state-space models outside the areas of actual activations (Figure~\ref
{figure1}). In addition, the two state-space solutions were
considerably smoother and in closer agreement with the simulated signal
than the MNE solution. The KF which computes its estimates based
exclusively on data up to the current time had a temporal lag with
respect to the true signal (Figure \ref{figure2}). That is, the
locations of the signal peaks and troughs were offset with respect to
their true temporal locations. In addition, the KF estimates
overestimated the true signal amplitude (Figure \ref{figure2}). The
backward pass of the FIS corrected both of these problems, yielding a
closer agreement with the true signal and a lower MSE (Figure \ref{figure2}).
Although in the analysis of the actual MEG experiment MNE and the
state-space models source estimates were in approximate agreement
(Figure \ref{figure5}), the magnitudes of the FIS estimates were more
consistent with those of the KF estimates, whereas the temporal
dynamics of the FIS estimates followed more closely those of the MNE
estimates (Figure \ref{figure5}). The KF estimates of the source
amplitudes were larger than those computed by the either MNE or the FIS
(Figure \ref{figure6}).
Although our findings establish the feasibility of using
high-dimensional state-space models for solving the MEG inverse
problem, they also suggest several extensions. In our analysis we
computed the KF and FIS estimates by testing different values of the
regularization parameter. This parameter can be estimated in an
empirical Bayesian [\citet{r35}] or a fully Bayesian framework. The
spatial component of the model can be improved by using the structure
of the particular experimental design to proposed specific forms of
the~$F$ state-transition matrix. There are a broad range of techniques that
have been used to accelerate computations in high-dimensional
state-space models. The forward model can be extended to include
subcortical sources. The MEG and EEG recordings are usually recorded
simultaneously. Our state-space paradigm can be applied to the problem
of estimating the sources from these two sources. These extensions will
be the topics of future reports.



\printaddresses

\end{document}